\RequirePackage{fix-cm}
\documentclass[smallextended]{svjour3}       
\smartqed  
\usepackage{amssymb}
\usepackage[pagewise,displaymath, mathlines]{lineno}
\usepackage{textcomp}
\usepackage{float}
\usepackage{bm}
\usepackage{graphicx,amsmath,amsfonts,bbm,setspace,url}
\usepackage{verbatim,mathrsfs}
\usepackage[mathcal]{eucal}
\usepackage{yfonts}

\usepackage{mathptmx}      
\usepackage{enumerate}
\usepackage[misc]{ifsym}
\usepackage{algorithm}
\usepackage{algpseudocode}

\usepackage[authoryear]{natbib}
\bibpunct{(}{)}{}{a}{}{;}

\begin{document}

\title{Nonrigid registration using Gaussian processes and local likelihood estimation
}

\author{Ashton Wiens \and
  William Kleiber \and
  Douglas Nychka \and
  Katherine R. Barnhart \and      
}

\institute{A. Wiens, (\Letter) \at
              Department of Applied Mathematics,
             University of Colorado, Boulder, CO. \\
            ORCiD: 0000-0002-7030-0602 \\
              Tel.: +316-680-4673 \\
              \email{ashton.wiens@colorado.edu} 
           \and
           W. Kleiber, \at
              Department of Applied Mathematics,
             University of Colorado, Boulder, CO. \\
             \and
             D. Nychka \at
             Department of Applied Mathematics and Statistics,
             Colorado School of Mines, Golden, CO. \\
              \and 
           K. Barnhart \at
              Cooperative Institute for Research in Environmental Sciences, University of Colorado, Boulder, CO \\
                Department of Geological Sciences, University of Colorado, Boulder, CO \\
                US Geological Survey, Landslide Hazards Program, 1711 Illinois St., Golden, CO, 80401
}

\date{Received: \;\;\;\; / Accepted: \;\;\;\;}
\maketitle

\begin{abstract}
Surface registration, the task of aligning several multidimensional point sets, is a necessary task in many scientific fields. 
In this work, a novel statistical approach is developed to solve the problem of nonrigid registration.
While the application of an affine transformation results in rigid registration, using a general nonlinear function to achieve nonrigid registration is necessary when the point sets require deformations that change over space.
 The use of a local likelihood-based approach using windowed Gaussian processes provides a flexible way to accurately estimate the nonrigid deformation. 
 This strategy also makes registration of massive data sets feasible by splitting the data into many subsets. 
 The estimation results yield spatially-varying local rigid registration parameters.
Gaussian process surface models are then fit to the parameter fields, allowing prediction of the transformation parameters at unestimated locations, specifically at observation locations in the unregistered data set. Applying these transformations results in a global, nonrigid registration.
A penalty on the transformation parameters is included in the likelihood objective function. Combined with smoothing of the local estimates from the surface models, the nonrigid registration model can prevent the problem of overfitting.
The efficacy of the nonrigid registration method is tested in two simulation studies, varying the number of windows and number of points, as well as the type of deformation.
The nonrigid method is applied to a pair of massive remote sensing elevation data sets exhibiting complex geological terrain, with improved accuracy and uncertainty quantification in a cross validation study versus two rigid registration methods.
\bigskip
\noindent


\end{abstract}

\section{Introduction}

Aligning or registering multidimensional point sets is a common problem arising in remote sensing, image and video processing, computer graphics, and computer vision. The goal in registration is to estimate a transformation which best aligns the moving data set with the target data set based on a measure of closeness. Rigid registration refers to the estimation and application of a single affine transformation to the coordinate system of the moving data set, whereas nonrigid registration involves a transformation function more general than an affine transformation. 

Classical approaches to registration rely on point or surface matching, or some variant thereof. Iterative closest point (ICP) (\citealt{icpworst2009}, \citealt{besl1992method}, \citealt{ezra2008}) and kernel correspondence methods \citep{tsin2004correlation} are common and have been developed in both the rigid and nonrigid settings. These algorithms iteratively assign correspondences between points (or surfaces) and then estimate a transformation which minimizes some distance function between the corresponding objects. Representing the points sets as mixtures of Gaussian densities is another strategy to solve the registration problem which reduces the computational complexity of the problem by avoiding discrete point matching, for example in \cite{jian2005robust} and \cite{zhou18}. The review articles by \cite{maiseli2017recent} and \cite{tam2013registration} detail many popular registration methods. \cite{tam2013registration} discriminate among the algorithms based on their optimization constraints and also how they assign correspondences; the majority of existing registration algorithms rely on a closest point criterion in point matching. 

Thin plate splines (TPS) and Gaussian processes (GP) play a prominent role in the parameterization of nonrigid deformations \citep{zhu2009nonrigid}. \cite{arad94} discuss two-dimensional radial basis function transformations, composed of an affine part and a radial function part. \cite{Bookstein} provides a similar mathematical development, however in the context of the physics of TPS as solutions to the biharmonic equation. \cite{Gilleland11} combines these ideas with climatological forecast verification, and \cite{chui03} combine the TPS parameterization with the robust point matching framework of \cite{gold1998new}. 

Motivated by the task of jointly registering and producing a surface estimate from two geologic data sets, \cite{Wiens2019} adapt methods from spatial statistics to produce a novel registration method based on maximum likelihood estimation. The algorithm embeds a rigid transformation within the Gaussian process regression maximum likelihood framework. The statistical nature of this method has the advantage of providing uncertainty measures about the estimated registration parameters. Although this method is not as fast as some traditional methods, results indicate this method can potentially provide a more accurate registration with uncertainty quantification, a feature not available in more algorithmic approaches. 

In this work, a novel nonrigid registration algorithm is developed based on a local likelihood estimation approach, expanding on the rigid registration model from \cite{Wiens2019}. The rigid registration likelihood procedure is applied using local windows of data, which gives insight into how these parameters change over the domain. Moreover, a surface model is fit to the estimated parameters, which allows the nonrigid transformation to be predicted at arbitrary locations. What is new in this work is the introduction of a global nonrigid deformation alignment model paired with an estimation approach that is scalable to massive data sets.

Both the rigid and nonrigid likelihood-based algorithms are fundamentally different from point matching and correspondence algorithms, as the points from both data sets are treated jointly as a realization from a stochastic process, so there is no point matching involved. On the other hand, the local likelihood estimation framework does share characteristics with some existing nonrigid registration methods. The methods that can handle data sets containing millions of points, including the nonrigid method developed here, often rely on downsampling the point clouds to reduce the computational burden \citep{Huang08}. 

Practical registration methods must have the capability to handle noise and outliers in the data sets. In point matching algorithms, it is common to prune correspondences which seem inconsistent. The likelihood-based methods are able to account for noise in the response variable within the statistical model, provided the parameters are correctly specified. Note that spatial models have been developed with the ability to account for location error (\citealt{cervone2015gaussian}, \citealt{cressie2003spatial}, \citealt{fanshawe2011spatial}, \citealt{fronterre2018geostatistical}); however, the errors in location are assumed to follow a distribution to be inferred, rather than viewing the problem as an inconsistency in the coordinate systems of the data sets. Finally, it is common to minimize a loss or energy function as the criterion used to estimate the deformation. In the method developed here, transformation parameters are estimated by minimization of a penalized likelihood function. The likelihood function follows from the standard Gaussian process framework, and the inclusion of penalties on the transformation parameters can prevent the intrinsic issue of overfitting when using nonrigid registration methods.

Local likelihood estimation for spatial data is not a new idea (\citealt{haas1990kriging}, \citealt{stein2018}, \citealt{nychka2018modeling}, \citealt{Wiens2020}). However, its application to nonrigid registration provides a completely new approach to this problem. Using local likelihood estimation for nonrigid registration has several advantages. It is amenable to the task of nonrigid registration because simpler local rigid registrations are estimated and pieced together. Furthermore, it gives flexibility in the estimation of the nonrigid transformation, being data driven and not constrained to the a priori choice of a nonlinear deformation function. Finally, local likelihood estimation also permits the use of parallel computing, significantly accelerating the computation of the registration. Parallelism, combined with subsampling, can allow massive data sets to be accurately registered at tractable computational cost.

In this work, the proposed nonrigid registration method is applied to a geomorphic data set collected via remote sensing. The pair of elevation point sets could be processed into a digital elevation model (DEM), but first require registration. The data were collected by piecing together several hundred photographs taken with a drone using a Structure from Motion (SfM) algorithm (\citealt{Fonstad:2013ja}, \citealt{James:2012fha}, \citealt{Westoby:2012bc}). With this form of processing, it is hypothesized that a nonrigid deformation may be necessary to align the geologic features seen in the data set.
 
In Sect. \ref{s:2}, the statistical model for nonrigid registration is presented. In Sect. \ref{s:3}, the estimation algorithm is outlined in the rigid and nonrigid cases, providing implementation guidelines. In Sect. \ref{s:4}, the efficacy of the nonrigid registration method is investigated in two simulation studies, testing how the type and severity of deformation affect estimation accuracy. How accuracy is affected by the number of windows used and the number of subsamples taken in each data window is also investigated. In Sect. \ref{s:5}, the nonrigid registration method is applied to a set of remote sensing scans with over sixty million points each, comparing to two rigid registration methods in a cross validation study.

\section{Statistical model}
\label{s:2}

The nonrigid approach developed in this work is a generalization of the rigid registration approach of \cite{Wiens2019} to a nonlinear setting. In the following, the general registration model which encompasses both the rigid and nonrigid registration cases is presented.

Let $Y_1(\mathbf s)$ be the data from the fixed point cloud and $Y_2(\mathbf s)$ be the moving point cloud to be registered, where $\mathbf s \in \mathbb R^2$ contains the x- and y- positional coordinates and $Y_i(\cdot)$ denotes the measured z-coordinate. The statistical process model for these data is
\begin{align} \label{s:gp}
  \begin{split}
    Y_1(\mathbf s) &= Z(\mathbf s) + \varepsilon_1(\mathbf s) \\
    Y_2(\mathbf s) &= \mu_z(\mathbf s) + Z(T(\mathbf s)) + \varepsilon_2(T(\mathbf s)),
  \end{split}
\end{align}
where $T$ is a transformation applied to the moving data set. 
Both datasets represent the same underlying continuous surface, $Z(\cdot)$, which is modeled as a mean zero Gaussian process. Specifically, a process model with a nonstationary Mat\'ern covariance function as in \cite{paciorek2006spatial} is adopted, with a smoothness fixed at unity. 
The measurement error or microscale variation $\varepsilon_i(\cdot)$ for $i=1,2$  are assumed to be independent mean zero Gaussian white noise processes with variance $\tau^2(\mathbf s)$. 

The model in Eq. \ref{s:gp} is quite general in allowing for a spatially-varying fixed mean function, $\mu_z(\mathbf s)$, white noise variance, $\tau^2(\mathbf s)$, and spatial registration function $T(\mathbf s)$. 
Some restrictions are required on the classes of functions entertained. In particular, it is proposed that these functions fall into a class of continuous and differentiable functions defined by realizations of a Gaussian process, specified below.

The full transformation is contained in the parameter functions $\mu_z : \mathbb R^2 \to \mathbb R$ and $T:\mathbb R^2\to\mathbb R^2$. $\mu_z$ represents the translation in the z-coordinate, and $T$ consists of a translation vector function $\mathbf r : \mathbb R^2 \to \mathbb R^2$ and a $2\times2$ orthogonal rotation matrix function $R : \mathbb R^2 \to \mathbb M^{2\times2}$ parameterized by an angle $\phi : \mathbb R^2 \to \mathbb R$ as follows.
\begin{align}
\label{eq:trnsf}
\begin{split}
  T(\mathbf s) &= R(\mathbf s) \mathbf s + \mathbf r(\mathbf s) \\  
    &= \begin{bmatrix}\cos \phi(\mathbf s) &\sin \phi(\mathbf s) \\-\sin \phi(\mathbf s) &\cos \phi(\mathbf s) \end{bmatrix} 
  \begin{bmatrix}s_x \\ s_y\end{bmatrix} 
  + \begin{bmatrix}r_x(\mathbf s) \\ r_y(\mathbf s) \end{bmatrix},
  \end{split}
\end{align}  
where $\mathbf s = [s_x, s_y]^T$ and $\mathbf r(\cdot)=[r_x(\cdot), r_y(\cdot)]^T$.  

Our specification is a nonrigid generalization of the model in \cite{Wiens2019} in which $\mu_z$, $\mathbf r$ and $\phi$ are assumed to be constant over the domain. 
The estimation of these parameters in the rigid and nonrigid cases are detailed in the Sect. \ref{s:3}.

\section{Estimation}
\label{s:3}

In the following, the estimation strategy is presented for both the rigid and nonrigid cases. The full nonrigid registration estimation algorithm is presented, followed by implementation details describing the options and tuning parameters available in the method.

In either the proposed nonrigid registration approach, or the rigid registration approach of \cite{Wiens2019}, the data under investigation are two point clouds whose spatial indices can be distinct. 
Let $\mathcal{U} = \left\{ (\mathbf s_{1,j}, Y_{1}(\mathbf s_{1,j})) \right\}_{j=1}^{n_u} $ and $\mathcal{V} = \left\{ (\mathbf s_{2,j}, Y_{2}(\mathbf s_{2,j})) \right\}_{j=1}^{n_v} $ be points sets whose elements are in $\mathbb{R}^3$. 
$\mathcal U$ will be called the target or fixed frame and $\mathcal V$ the moving frame; the choice of $Y_1$ being the target frame and $Y_2$ being the moving frame is inconsequential as the model in Eq. \ref{s:gp} can be rewritten to accommodate the opposite setup with an analogous parameterization. 
Let $\mathbf Y_1 = (Y_{1}(\mathbf s_{1,1}), \cdots, Y_{1}(\mathbf s_{1,n_u}))$, $\mathbf Y_2 = (Y_{2}(\mathbf s_{2,1}), \cdots, Y_{2}(\mathbf s_{2,n_v}))$, and $\mathbf Y = (\mathbf Y_1^T,\mathbf Y_2^T)^T$. 
Analogously define the continuous Gaussian process components as $\mathbf Z_1, \mathbf Z_2$ and $\mathbf Z$.

\subsection{Rigid registration estimation}
\label{ss:rigid}

In the rigid registration algorithm, global estimation is performed to infer all mean and covariance parameters, including transformation parameters, using all available data $\mathbf Y$ simultaneously. 
Concatenating all parameters to be estimated into a vector $\bm \theta$, our estimator is $\hat{\bm \theta} = {\rm argmin}_{\bm \theta} f(\mathbf Y, \bm \theta)$ where 
\begin{align}  \label{eq:l.plus.p}
  f(\mathbf Y, \bm \theta) = 
  {\cal L}(\mathbf Y,\bm \theta) + {\cal P}(\bm \theta).
\end{align}
Here, ${\cal L}(\mathbf Y,\bm \theta)$ is a negative log likelihood, while ${\cal P}(\bm \theta)$ is an optional penalty term that serves to regularize estimates of $\bm \theta$.

Assuming $\mathbf Y$ is a realization from a stationary Gaussian process with Mat\'ern covariance, ${\cal L}$ is multivariate normal negative log likelihood with mean $\mathbf m$ and covariance matrix $\Sigma + \tau^2 I$.  For reference, the Mat\'ern covariance function is parameterized in this work as $\operatorname{Cov}(Z(\mathbf s), Z(\mathbf s')) = \sigma^2\frac{2^{1-\nu}}{\Gamma(\nu)} \left(\frac{d}{a} \right)^{\nu} K_{\nu}\left(\frac{d}{a}\right)$, where $d=\|\mathbf s-\mathbf s'\|$ is the Euclidean distance between points, $\sigma^2$ is the marginal variance, $a>0$ a spatial range parameter and $\nu>0$ the smoothness.  $\Gamma$ is the gamma function, and $K_{\nu}$ is the modified Bessel function of the second kind of order $\nu$.
Given the model in Eq. \ref{s:gp},  the mean of $\mathbf Y$ is $\mathbf m = (\mathbf{0}_{n_u}^T, \,\,\mu \mathbf{1}_{n_v}^T)^T$, where $\mathbf{0}_{n_u}$ and $\mathbf{1}_{n_v}$ are zero and unit vectors of lengths $n_u$ and $n_v$, respectively. 
The covariance matrix for $\mathbf Y$ is 
\[
  \Sigma + \tau^2 I =  
  \begin{bmatrix}
    \operatorname{Cov}(\mathbf Z_1,\mathbf Z_1) & \operatorname{Cov}(\mathbf Z_1,\mathbf Z_2) \\
    \operatorname{Cov}(\mathbf Z_2,\mathbf Z_1) & \operatorname{Cov}(\mathbf Z_2,\mathbf Z_2)
  \end{bmatrix}
  + 
  \tau^2 I \, ,
\]
where $I$ is an identity matrix of dimension $n_u+n_v$. 
One key insight is that the cross covariances $\operatorname{Cov}(\mathbf Z_1,\mathbf Z_2),\operatorname{Cov}(\mathbf Z_2,\mathbf Z_1)$ and the covariance matrix of the second data set only $\operatorname{Cov}(\mathbf Z_2,\mathbf Z_2)$ involve the rigid transformation function $T$, see \cite{Wiens2019} for details.

The motivation for including a penalty component is that during initial explorations the optimization would sometimes allow for the translation and rotation parameters to take on very large values, effectively making the supports of $\mathbf Y_1$ and $\mathbf Y_2$ disjoint, while it is expected that their optimal values for alignment are smaller. 
Thus, a penalty term of the form
\begin{align}  \label{eq:penalty}
  {\cal P}(\bm \theta) = 0.5 \lambda (r_x^2 + r_y^2 + \mu_z^2) + 
   \log ( I_0(\kappa)) - \kappa \cos \phi
\end{align} 
is used, where $I_0$ is the modified Bessel function of order 0, $\lambda$ is the tuning parameter for the penalty on the translation parameters and $\kappa$ is the tuning parameter for the penalty on the rotation parameter $\phi$. 
In the Bayesian context, these penalties are equivalent to placing Gaussian and von Mises distribution priors on the translation and rotation parameters, respectively. 
The penalty terms keep the magnitude of the registration parameters small, and additionally box constraints can be placed on these parameters in the optimization.

Minimizing the objective function $f(\mathbf Y,\bm \theta)$ results in estimates for the transformation parameters $\bm \theta$, which include $r_x, r_y, \mu_z$, and $\phi$, as well as $\tau$ and the covariance parameters specifying $\Sigma$ (e.g., the process variance and correlation range).

\subsection{Nonrigid registration estimation}

Estimation of the nonrigid registration model in Eq. \ref{s:gp} can be envisioned in many ways: one possibility would be to take a Bayesian approach, putting, say, Gaussian process priors on the parameter functions, followed by sampling their posterior distributions. 
Another approach would be to explicitly parameterize these surfaces as a linear combination of spatial basis functions. 
However, both of these approaches are not feasible for our setup with tens of millions of data points per point cloud.  
A local likelihood approach to estimation is thus introduced that allows for computation on very high dimensional data sets. 

To briefly outline the estimation approach, the spatial domain is divided using a moving window, the parameters are locally estimated using the rigid registration estimation strategy given in Sect. \ref{ss:rigid} within that window using a subset of the data, and finally local estimates are aggregated using a spatial surface model. 
More concretely, at any given moving window center, the data are subsampled and then used to fit the rigid registration model given in Sect. \ref{ss:rigid}, essentially assuming the registration parameters are locally constant, and the process covariance function is locally stationary. 
The same isotropic Mat\'ern covariance model is used as in the rigid setting during local estimation and prediction. 
These local estimates are then used as samples of the parameters $\bm \theta$, to which independent spatial surface models are fit.
The full approach is described in Algorithm \ref{alg}.

\begin{algorithm}[!ht]
\caption{Gaussian process nonrigid registration}
\label{alg}
\begin{algorithmic}[1]
\State Set up grid $\mathcal{G} = \left\{ \mathbf g_k \right\}_{k=1}^{n_g}$ to locate the centers of overlapping windows $\left\{ \mathcal B_k\right\}_{k=1}^{n_g}$, and partition the data using these windows yielding $\mathcal D^U= \left\{ \mathcal D^U_k \right\}_{k=1}^{n_g} $, where each $\mathcal D^U_k$ contains the observations in $\mathcal U$ that lie within $\mathcal B_k$. Define $\mathcal D^V$ and $\mathcal D^V_k$ similarly. 
\State For the data set $\mathcal U$, subsample $N$ points from each window $\mathcal D^U_k$, resulting in $\mathcal{\tilde{D}}^U_k$ (and similarly for $\mathcal V$).
\State Perform Gaussian process embedded rigid registration using data  $\mathcal{\tilde{D}}^U_k \cup \mathcal{\tilde{D}}^V_k$ for each window $\mathcal B_k$, yielding local estimates for $\hat{\bm{\theta}}_k \supseteq \{ \hat{r}_{x,k}, \hat{r}_{y,k}, \hat{\mu}_{z,k}, \hat{\phi}_k, \hat{\tau}_k, \hat{\Sigma}_k \}$ for $k = 1, \cdots, n_g$. Assign the estimated parameters to the centers of the estimation windows $\mathbf g_k$ for $k = 1, \cdots, n_g$. 
\State For each univariate parameter in the vector $\hat{\bm{\theta}}_k$, independently fit a predictive spatial surface model using the estimated spatially varying transformation parameters $\hat{\bm{\theta}}_k$ as observations. 
\State Using these fitted models, predict the transformation parameters at all of the observation locations in the moving data set  $\{ \mathbf s_{2,j} \}_{j=1}^{n_v}$. Denote the predictions $\{ \bm{\theta}^*(\mathbf s_{2,j}) \}_{j=1}^{n_v}$, which is a set of vectors of parameters, each vector of length $n_v$. In particular, predict the transformation parameters $r^*_x(\mathbf s_{2,j}), r^*_y(\mathbf s_{2,j}), \mu^*_z(\mathbf s_{2,j}), $ and $ \phi^*(\mathbf s_{2,j})$, (thus implying $\bm r^*(\mathbf s_{2,j})$ and $R^*(\mathbf s_{2,j})$), for $j = 1, \cdots, n_v$.
\State Apply the predicted transformation parameters to the moving data set, resulting in a set $\mathcal{\bar{V}} = \left \{ (\mathbf{s}^{*} _{2,j}, Y^{*} _{2,j}) | \mathbf{s}_{2,j}^{*} = R^*(\mathbf s_{2,j}) \mathbf s_{2,j} + \bm{r}^*( \mathbf s_{2,j}) \textrm{ and } Y^{*} _{2,j} = Y_{2,j} + \mu^*_z(\mathbf s_{2,j}) \right \}_{j=1}^{n_v}$, which is registered to the coordinate system of $\mathcal{U}$.
\end{algorithmic}
\end{algorithm}

\subsection{Implementation details} 

Algorithm \ref{alg} will now be discussed in more detail, explaining the modeling choices involved in each step. 
Some choices track this paper's applications, but might be modified for other registration problems. 
Five of the most important parameters that need to be chosen are the number of estimation locations $n_g$, the size of windows, and the overlap among windows; the number of subsamples $N$ used in each window for the local estimation; the penalty type and magnitude of the tuning parameters (e.g., $\lambda$ and $\kappa$), and optimization parameter constraints; the covariance function for the Gaussian process $Z(\cdot)$ in which the rigid transformation is embedded, defining $\Sigma_i$; and finally the predictive spatial surface model to fit to the locally estimated parameters.

Many windows can be used to assure the signal is captured in the spatially-varying rigid registration parameters. However, using too many windows increases the computational burden. Most importantly, the window size in the local estimation procedure must be chosen to contain features present in both data sets and so allow the local rigid registration Gaussian process model to identify and estimate the rigid registration. A simulation study is conducted in the following section to investigate this issue.
Overall, one should choose the number of windows and overlap so that the signal of the deformation is observed in the raw estimates before a spatial model is fit. 
In this work, local windows overlap significantly (50 to 75$\%$ overlap between adjacent windows) so that the estimated parameters result in smoothly varying spatial fields.

For spatially dense point clouds, the number of points in each window may still be too large to make fitting a Gaussian process quickly feasible. Therefore, one can subsample points from both data sets in each window to decrease the computational burden and still end up with accurate results. Some windows which include near-vertical terrain (e.g., a cliff in observations of terrain) will contain many more points overall than the average window. In this case, a lower sampling proportion of points can be taken for the likelihood estimation, or more windows can be used in this region (i.e., windows are more closely spaced together). This adaptive strategy is not explored further in this work.

Some device must be used in order to force overlap between the two subsets of data within each window; otherwise, there is no incentive for the algorithm to intersect the two data sets during alignment. Tight constraints on the translation parameters are used in \cite{Wiens2019} based on empirical observation of the magnitude of the transformation that was needed. In this work, the use of a penalty term on the transformation parameters is utilized. The penalty incentivizes the two subsets of data in each window to overlap without tight box constraints on the transformation parameters. If the penalty is too large, the transformation parameters will be estimated near zero. One can gain insight by carrying out the local likelihood estimation procedure without the penalty parameter. The magnitudes of the estimated transformation parameters combined with the value of the likelihood function at a minimum with the penalty parameter set at 0 can inform how to choose the size of the penalty parameter in subsequent application of the method. 

In each local rigid registration, the data sets are modeled using a Gaussian process with Mat\'ern covariance, following \cite{Wiens2019}, due to its flexibility. However, the rigid registration could be embedded in another covariance model with ease if the data were assumed to follow a specific distribution.

Finally, to piece together the local rigid registration parameters into a surface, either a Gaussian process or thin plate spline is used.
Both choices provide flexible spatial models allowing for prediction of the parameters at unobserved locations. The nonrigid method produces stable local estimates that vary smoothly over space, if one includes some thought in determining subsample size and penalties on parameters. Either model is able to separate the signal and noise components in these estimates and predict accurately at unestimated locations. Thin plate splines are used in the simulation studies. On the other hand, in the data analysis a Gaussian process with nonstationary covariance structure is used, provided in the \textsc{LatticeKrig} package \citep{LK} in \textsc{R} \citep{Rcore}. The default settings for this Gaussian process model approximate a thin plate spline but scale to large data sets. Additionally, this model is capable of providing uncertainty estimates about predictions and produce conditional simulations, which enable us to calculate forecast scores such as the continuous rank probability score (CRPS).

\section{Simulation Studies}
\label{s:4}

\subsection{Types of deformations}

To test the efficacy of the proposed nonlinear registration algorithm, two simulation studies are implemented designed illustrate when the method succeeds and fails. 
The first simulation study investigates how the type of deformation and which parameters the deformation is applied to affects the estimation accuracy. 
The second study explores the interplay between the number of estimation locations and the number of subsamples used for each local estimation. 

For both studies, the same general setup is used. A spatial data set is simulated using a Gaussian process with zero mean and Mat\'ern covariance function with parameters $a = 3, \nu = 1, \sigma^2 = 2.5, \tau^2 = 0.0001$. The observation locations for both data sets are uniformly sampled within a $[0,6]\times[0,6]$ square. The data set is split in half, and one half is de-registered with a known deformation. Then, the nonrigid registration algorithm is used to recover the spatially-varying registration parameters. 
In all cases, 10,000 spatial points are generated, split in half so each data set contains 5,000 points.
 
The first simulation study attempts to answer what kind of deformations the method is capable of estimating. Two different strategies are used to generate the true deformations that de-register the second data set: a simple quadratic field, and a random realization from a Gaussian Mat\'ern process. The Mat\'ern deformations are drawn from a Gaussian process with zero mean and Mat\'ern covariance with parameters $a = 6, \nu = 2, \sigma^2 = 0.01, \tau^2 = 0.00001$. One Mat\'ern field is drawn for each translation and rotation parameter, and then applied to the moving data set based on the design of the study.
The quadratic deformation for the x-coordinate, for example, is defined as follows.
$$ m_x = \alpha_1 s_x^2 + \alpha_2 s_y^2 + \beta_1 s_x + \beta_2 s_y,$$
where $\alpha_{1,2}$ are drawn from a uniform distribution $ U(-0.002, 0.002)$, and $\beta_{1,2}$ are drawn from a uniform distribution $ U(-0.04, 0.04)$. The transformation $m_x$ is then applied to the x-coordinate in the moving data set. The same equation governs how the translations in the y- and z-coordinate $m_y$ and $m_z$ are generated, as well as the rotation $m_\phi$.

To investigate how the complexity of the deformation affects estimation accuracy, the Mat\'ern and quadratic deformations are applied to just single parameters as well as groups of parameters. There are five cases: the deformation is applied to the x-coordinate, the z-coordinate, the rotation angle $\phi$, the three translation coordinates (x, y, and z), and all four parameters (x, y, z, and rotation angle $\phi$).

\begin{figure*}[t]
\centering\includegraphics[width=0.9\linewidth, height=300px]{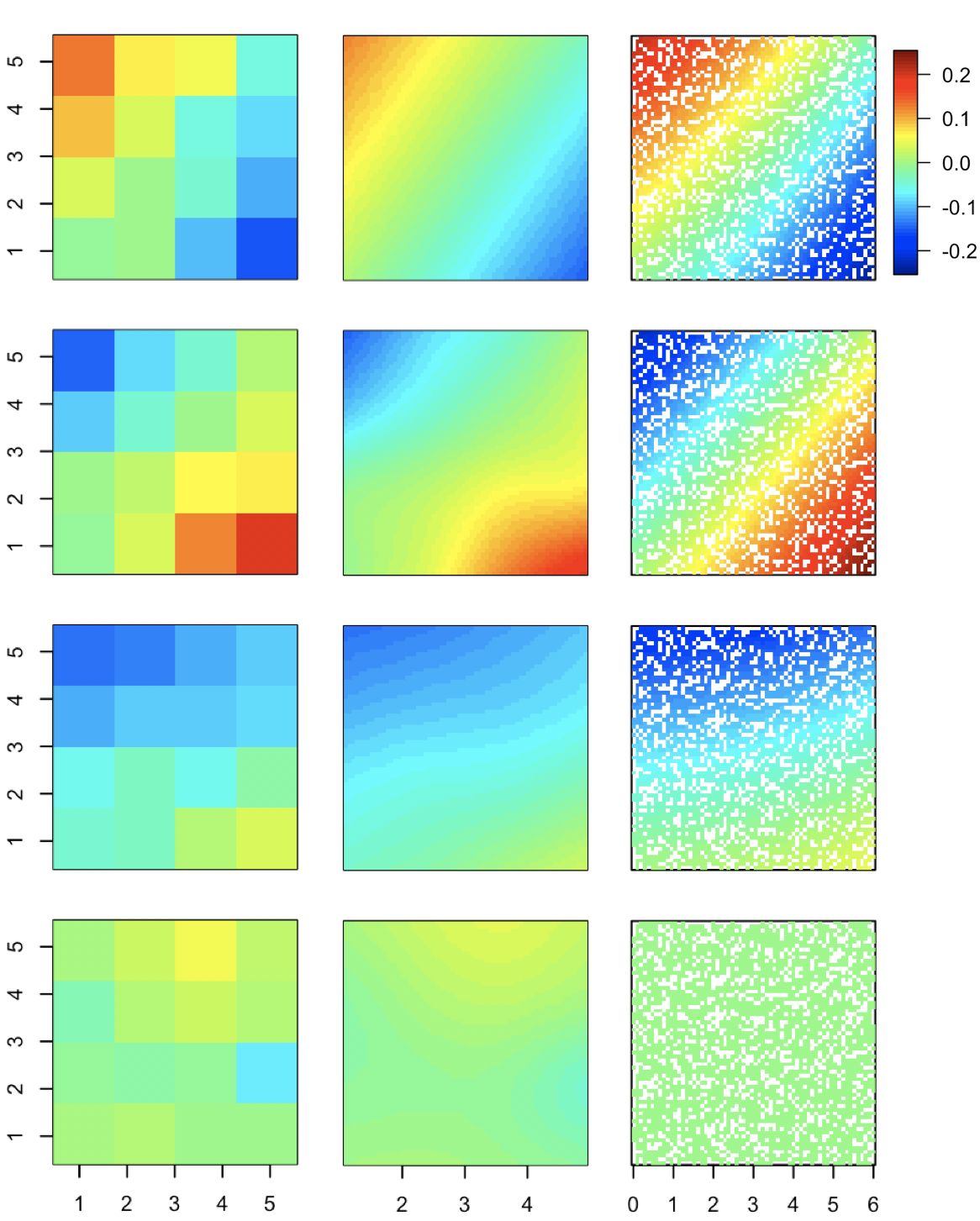}
\caption{ The first column shows the estimated transformation parameters, the second column shows the thin plate spline surface predictions fitted to these data, and the third column shows the true deformation parameters which are applied to de-register the moving data set}
\label{f:ss}
\end{figure*}

To estimate the registration parameters, a $4\times 4$ grid of windows with 50\% overlap between adjacent windows is used. The $4\times4$ grid is shown in the first column of Fig. \ref{f:ss}. 100 data points are subsampled within each window, and the penalty tuning parameters on the translation parameters and rotation parameter are set at 5 and 100, respectively. Box constrains are set on the translation parameters of $\left(-1,1\right)$ and on the rotation parameter of $\left(-\pi/4 + 0.1, \pi/4\right)$ in the optimization. These settings reflect those used in the data analysis example.
Finally, a thin plate spline is fit with default settings using the \textsc{Tps} function in the \textsc{fields} package \citep{fields} in \textsc{R} \citep{Rcore}.

In all cases, to quantify the recovery of the nonlinear registration, the estimated registration parameters, denoted with a hat as in $\hat{m}_x$, are compared to the true registration parameters, for example $m_x$. 
Figure ~\ref{f:ss} shows an example simulation: the true quadratic deformations are shown in the third column and the raw estimates and fitted surfaces are shown in the first and second columns, respectively.

The results of the simulation study are summarized in Table ~\ref{t:1}. The experiment is executed $n=30$ times and the average normalized root mean squared error (NRMSE) is computed for each cell in the table. The root mean squared error (RMSE) is normalized by the mean deformation over all coordinates with nonzero deformation. For example, when the deformation is applied only to the x-coordinate, the NRMSE for the x-coordinate is calculated as follows. Let $r_x = (m_x - \hat{m}_x)^2$ be the vector of squared errors in the x-coordinate. Then
$$\textrm{NRMSE}_x = \sqrt{\bar{r}_x} / \bar{m}_x,$$
where the bar denotes the sample mean of a vector.
When the deformation is applied to all coordinates, the NRMSE for the x-coordinate is
$$\textrm{NRMSE}_x = \sqrt{\bar{r}_x} / \bar{m},$$
where the vector $ m = (m_x, m_y, m_z, m_\phi )$ is the concatenation of the deformations applied to each data point for all coordinates. In this way, the RMSE is normalized using the mean of all deformations that are actually applied.

These results indicate that deformations of both types are recovered effectively. Interestingly, a quadratic deformation of the type that is applied here could possibly be estimated using a rigid registration with a scaling/stretch parameter. However, this study indicates that this deformation can also be effectively estimated using piecewise rigid transformations.

It appears that the nonlinear registration algorithm is able to well-estimate deformations applied to the x-,y-, and z-coordinates, but has more trouble recovering deformations involving a rotation. However, the rotation angle is only inaccurately estimated if an angular deformation is introduced. If a deformation is applied to the rotation angle, this also introduces error in estimating the x-, y-, and z-coordinates. Another insight from Table \ref{t:1} is that when the true parameter is zero (i.e., no misalignment or rotational misalignment) then the estimated parameters are particularly accurate, suggesting that the approach tends to not introduce deformations when they are not needed.



\subsection{Windowing and subsampling}

In the second simulation study, the effect of the number of estimation locations and the number of subsamples of the estimators is investigated.

Grids of sizes $3\times 3, 4\times 4$, and $5\times 5$ are used to locate the centers for overlapping windows. Subsamples are taken within each window of size 25, 50, 100, and 200. These experimental settings kept the computation reasonable by limiting both the number of local estimation tasks and also the total number of observations in each Gaussian process estimation. 

The moving data set is deformed using a realization from a Gaussian process with Mat\'ern covariance, as in the first simulation study. The covariance parameters used to generate the deformation are known and fixed in the estimation. 
Furthermore, only the x-coordinate is deformed and estimated, meaning there is only one registration parameter to estimate ($\phi$ and the y- and z-coordinate translation parameters are fixed at 0). The box constraint on the x-coordinate is set at (-1,1) in the optimization. The overlap between adjacent windows and the regularization tuning parameters are identical to the previous simulation study, as well as the thin plate spline surface fits with default settings.

This experiment is repeated $n=30$ times to compute average NRMSE, which is shown in Table \ref{t:2}. In this case, the deformation is only applied to the x-coordinate, so the RMSE is normalized as follows.
$$\textrm{NRMSE}_x = \sqrt{\bar{r}_x} / \bar{m}_x.$$

The trend is clear from these results: the nonlinear registration algorithm is more accurate when using more windows and more subsamples. However, since the task becomes much more computationally expensive when using more subsamples and more windows, a balance must be found based on how the registration parameters vary over space. If there is not much variance in the registration parameters over space based on exploratory analysis, few windows and subsamples are needed, such as in this example. If the point clouds exhibit highly complex variation in the response variable, more subsamples and windows may be required to obtain accurate results.


\section{Drone-based point clouds from the Chalk Cliffs}
\label{s:5}

The proposed nonrigid Gaussian process registration method is applied to a pair of geomorphic elevation data sets with over 100 million points. The data are point sets processed from drone-based remote sensing scans of the Chalk Cliffs debris flow basin in Colorado. The Chalk Cliffs receives debris from a 0.37 km$^2$ watershed at the base of Mount Princeton in the Arkansas River Valley. The considered data set results from two flyovers of the fan during the same day, 10/16/2017, at 40 meter flight altitude. No large scale erosion or deposition occurred between flights. \cite{barnhart2019} and \cite{Wiens2019} provide details about the geologic motivation. It is hypothesized that due to the SfM algorithm, a nonrigid registration may be necessary for proper alignment.

\begin{figure*}[t]
\centering
\includegraphics[width=1\linewidth]{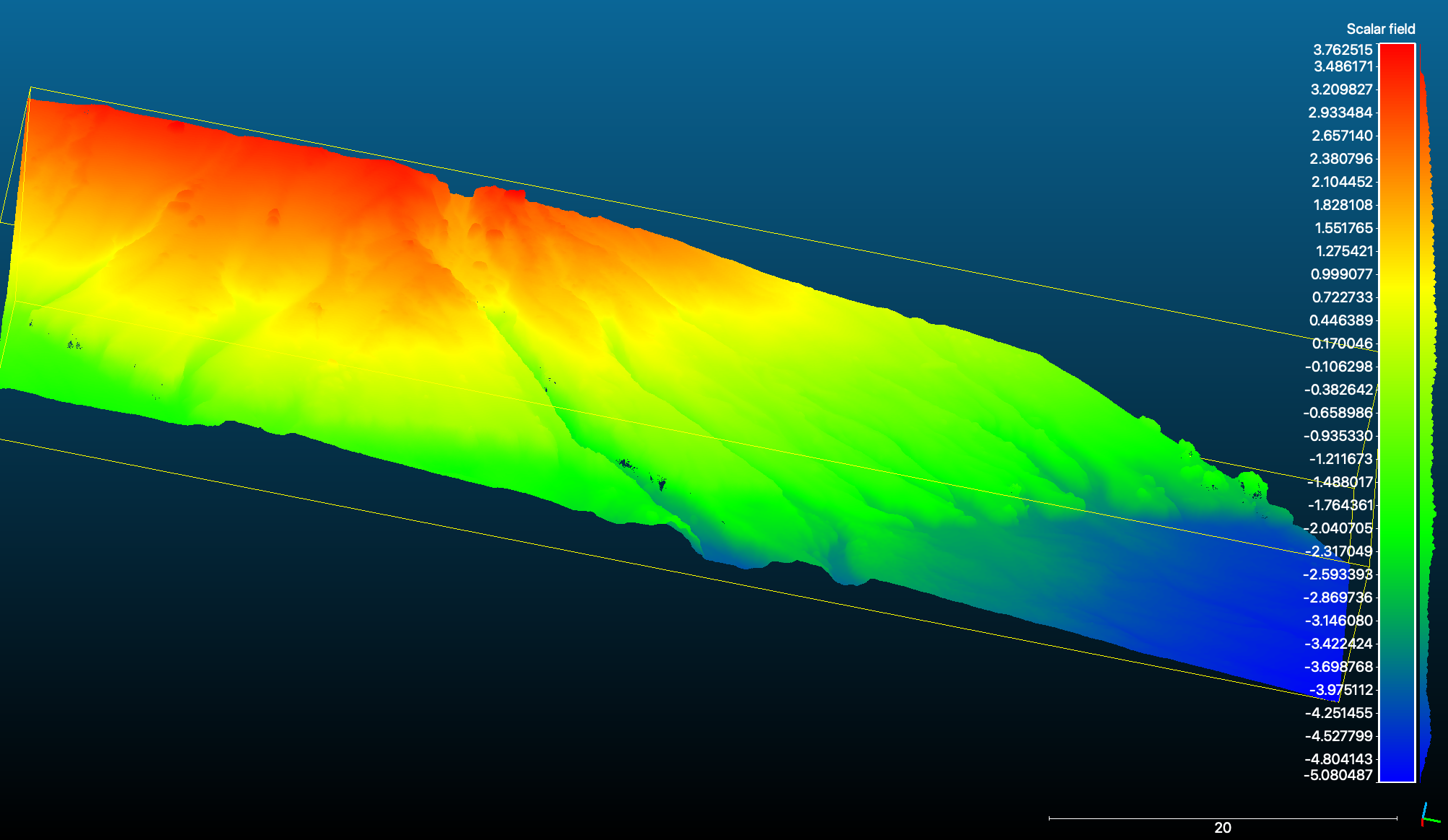}
\caption{One scan showing the full data set. This is the target data set containing 6,775,704 points in total. The units displayed in the scales are meters}
\label{f:1}
\end{figure*}

To streamline the application for this work, an area is identified using similar geographic features in both scenes, and then the data are cropped manually, resulting in areas whose spatial boundaries are close but not identical. In total, there are around 60 million data locations in each cropped scan of the region; one scan is shown in Fig. \ref{f:1}. 

\begin{figure*}[t]
\makebox[\textwidth][c]{\centering\includegraphics[width=340px]{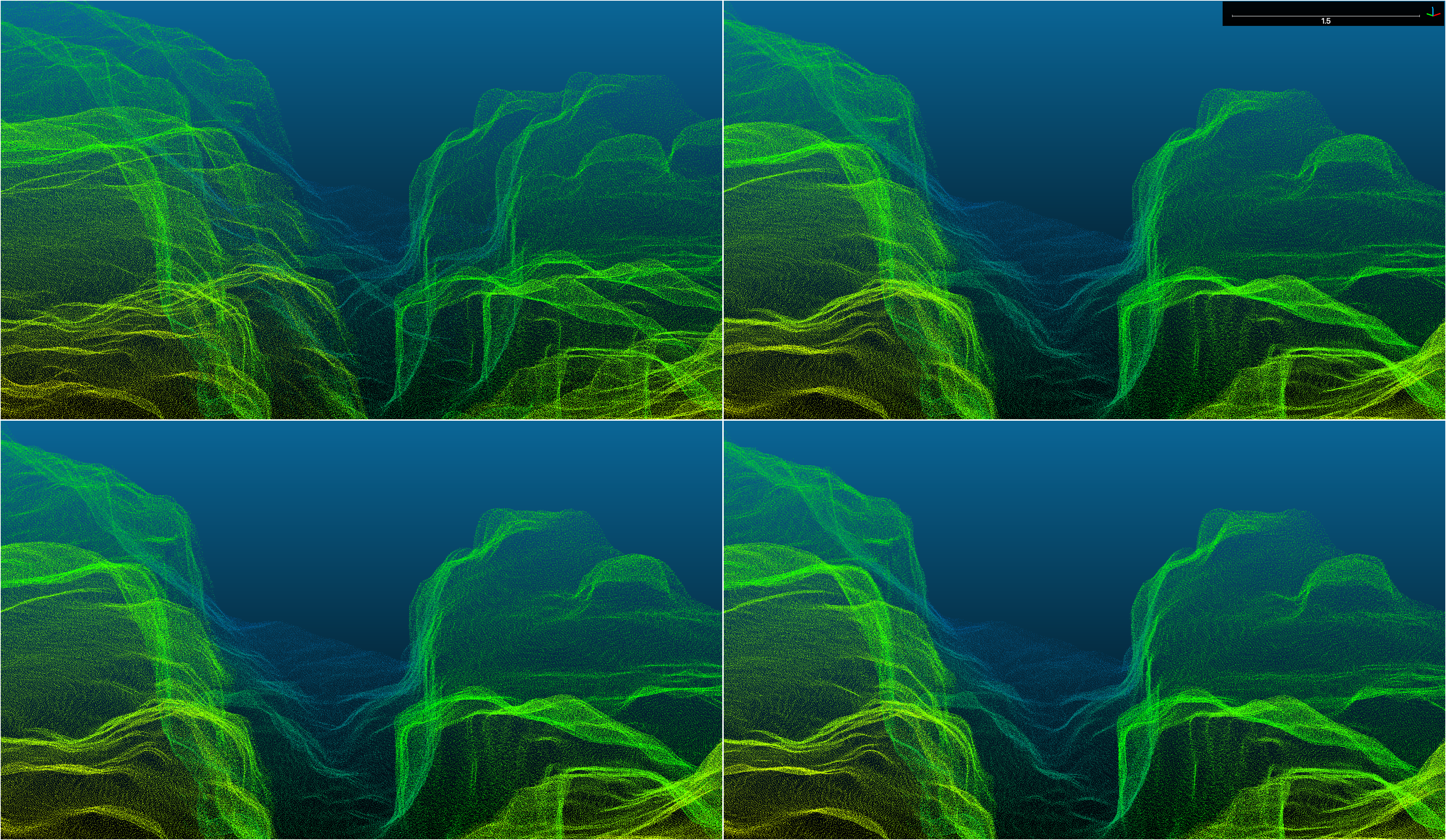}}
\caption{The top row shows both data sets before and after the nonrigid registration is applied, zoomed in centered on a ravine. The bottom row shows two sample registrations based on conditional simulations of the registration parameters. The colorbar scale is identical to Fig. \ref{f:2}}
\label{f:2}
\end{figure*}

The nonrigid registration method developed in this paper is compared with the likelihood-based rigid registration method, as well as the iterative closest point algorithm. The predictive quality of the methods is assessed in a cross validation study, holding out $0.1\%$ of the points from the fixed data set to be used as the testing set, and the remaining observations in the fixed data set and the entire moving data set are used to estimate the registration parameters. ICP is performed in the open source CloudCompare software, version 2.9.1, and to generate Figs. ~\ref{f:1} and ~\ref{f:2}. To match the registration capabilities of the rigid and nonrigid likelihood-based algorithms, the rotation is restricted to the $xy$-plane when using ICP.

For the nonrigid registration, a $4\times 12$ grid of windows is used with 66\% overlap. 250 subsamples from each data set are used in the local estimation, and the regularization tuning parameters are fixed at $\lambda=5$ and $\kappa=100$. The box constraints in the optimization are set at $\left(-1.2,.1.2\right)$ on the translation parameters and $\left(-\pi/4 + 0.1, \pi/4\right)$ on the rotation parameter. The rotation angle constraints are unrestrictive, and the translation parameter constraints encompass the true values, so in this case are also unrestrictive, but help reduce the search space and prevent complete non-overlap between the data sets. The locally estimated rigid registration parameters are fit with \textsc{LatticeKrig} Gaussian process models.

Since the data sets are too large to perform standard kriging, a local kriging procedure is used \citep{haas1990kriging, haas1990lognormal, haas95} to predict the response at the locations of the held-out observations for all registration methods. In particular, the 1000 closest points to the observation location to be predicted are used, and with these points standard kriging is performed using the stationary Mat\'ern covariance function. The covariance parameters used in the kriging prediction are either constant for the two rigid registration methods, or spatially varying as estimated in the nonrigid registration case, detailed below.

The covariance parameters for the rigid registration method are estimated as part of the rigid likelihood-based procedure. Since the full data sets cannot be used, 500 observations from each data set are subsampled and used to fit the rigid registration model. 
The transformation parameters are applied to the moving data set, and the covariance parameters are used to perform the local kriging. Similarly, for ICP, after the rigid registration parameters are applied to the moving data, the covariance parameters are estimated using 500 observations from both data sets, then used in the local kriging procedure. 

For the nonrigid registration, spatially varying covariance parameters are estimated along with the transformation parameters. Similar to the transformation parameters, the estimated covariance parameter fields are smoothed using \textsc{LatticeKrig} Gaussian process models, and then used to predict the covariance parameters at the locations of the observations in the testing set. The \textsc{LatticeKrig} model makes prediction at the large number of locations in the testing set tractable.


The cross validation results are shown in Table \ref{t:rmse}. In addition to calculating RMSE using the true and predicted values of the response for the testing set, CRPS is calculated to assess the accuracy of the uncertainty quantification. To calculate the variance of the predictions, a simulative approach is used. For all three methods, CRPS is calculated using 30 simulations. 

For the nonrigid method, the locally-estimated rigid registration parameters have uncertainties approximated in the optimization by the numerical Hessian. Using these uncertainties, registration parameters are simulated at the locations of the centers of the moving windows. A \textsc{LatticeKrig} model is fit to these simulated parameters, and then a conditional simulation is produced at the observation locations of the moving data set, and then applied. This technique propagates the two levels of uncertainty into the registration, from the local estimation and the predictive surface model fitting. This procedure was repeated 30 times and these empirical simulations of the predictions at the testing set locations were used to calculate the CRPS.

For the rigid likelihood-based method, the rigid registration parameters are simulated using the uncertainties from the numerical approximation to the Hessian, and then applied to the entire moving data set. Then local kriging is performed using simulations of the Mat\'ern covariance parameters at the testing set locations, once again from the approximate distribution given by the numerical Hessian.
For ICP, only the covariance parameters are simulated prior to local kriging as just described. Specifically, rigid registration using ICP produces no uncertainty measures, so the parameter estimates are treated as fixed.

Overall, the nonrigid registration algorithm achieves lower RMSE and CRPS than the two rigid registration methods, indicating superior accuracy of the registration and also quantification of the uncertainty. Moreover, visually the two data sets are more accurately aligned as seen in the top row of Fig. ~\ref{f:2}, which shows a close-up view of both data sets before and after registration using the nonrigid registration method in the left and right panels, respectively. There is still clear misalignment, albeit minor. The fact that the approach is statistical is critical here. While there may be a best alignment, it is acknowledged that the estimated nonrigid registration is not perfect; however, the uncertainty about the registration can be quantified. 

The bottom row of Fig. \ref{f:2} displays two registrations incorporating uncertainty using the simulative technique used in calculating CRPS. The bottom right panel contrasts a more accurate registration with a less accurate registration in the bottom left panel. Visually this shows how the uncertainty estimates encompass the true nonrigid deformation.


\section{Discussion}

The nonrigid registration algorithm developed in this paper provides an extremely flexible method for registering complex data sets. Moreover, the local likelihood estimation within each window can be treated as an independent task, and can thus utilize parallel processing. For these reasons, this method is capable of handling massive data sets. 

When using the nonrigid registration method, several modeling choices must be leveraged to obtain accurate results. One must balance computational demands with model flexibility. The choice of the number of windows and the number of subsamples used in each window in the estimation procedure influences the degree to which the registration parameters are allowed to vary over space, and also controls the computational burden of the algorithm. If a highly nonrigid deformation is required, many windows should be used. If the surface is complex or noisy, many subsamples should be used in each window to ensure proper local rigid registration.

Overfitting is intrinsically associated with nonrigid registration methods, and regularization is essential to avoid this problem \citep{tam2013registration}. The method developed in this paper has the capability for accurate and flexible nonrigid registration, but it also suffers from the possibility of overfitting. If too many windows are used in conjunction with too small of a penalty parameter, the data can be erroneously registered. However, the inclusion of a penalty parameter and the smoothing of the spatial surface model help combat overfitting associated with nonrigid registration methods. While the penalty parameter is advantageous in this regard, it can also influence the estimation of the registration parameters if it is too large. 

The choice of predictive spatial surface model can affect the final results. 
 Both Gaussian processes and thin plate splines provide flexible models which allow for effective smoothing of the locally estimated rigid registration fields. One could use a stationary or nonstationary covariance function in the Gaussian process model based on the context, or a thin plate spline is viable if uncertainty quantification is not important in the analysis.

The likelihood-based nonrigid registration method is a unique and novel statistical approach to solving this problem which can help further the field of registration. Through two simulation studies, it has been shown that the method can effectively recover deformations of different types, and the effect of using different tuning parameters in the method was explored. Finally, in the data analysis example, more accurate predictions and uncertainties were achieved in a cross validation study using the nonrigid method compared with two rigid registration approaches.

\section{Acknowledgements}
This work was supported by the National Science Foundation (awards DMS-1811294 and DMS-1923062). Additionally, Barnhart was supported by NSF award EAR-1725774. This work utilized resources from the University of Colorado Boulder Research Computing Group, which is supported by the National Science Foundation (awards ACI-1532235 and ACI-1532236), the University of Colorado Boulder, and Colorado State University \citep{rmacc}.

\section{Tables and figures}

%


\begin{table}[ht]
\centering
  \caption{NRMSE of estimating each deformation parameter for varying deformation types, normalized by the mean deformatia all coordinates with nonzero deformation}
   \label{t:1}
\begin{tabular}{rr|rrrr}
   \hline
  \textbf{Deformation type} &  \textbf{Applied to} &  $\operatorname{NRMSE}_x$ &   $\operatorname{NRMSE}_y$ &   $\operatorname{NRMSE}_z $ &   $\operatorname{NRMSE}_\phi $ \\ 
    \hline
 Quadratic & $x$ & 0.114 & 0.069 & 0.076 & 0.169 \\  
 Quadratic & $z$ & 0.109 & 0.102 & 0.105 & 0.124 \\  
Quadratic & $\phi$ & 3.956 & 3.951 & 0.421 & 0.711 \\  
Quadratic & $xyz$ & 0.132 & 0.119 & 0.105 & 0.196 \\  
Quadratic & $xyz\phi$ & 3.620 & 3.253 & 0.343 & 0.670 \\ 
Matern & $x$ & 0.121 & 0.080 & 0.075 & 0.108 \\  
Matern & $z$ & 0.083 & 0.080 & 0.135 & 0.074 \\  
Matern & $\phi$ & 3.618 & 3.634 & 0.206 & 0.571 \\  
 Matern & $xyz$ & 0.129 & 0.118 & 0.131 & 0.142 \\  
 Matern & $xyz\phi$ & 3.150 & 3.091 & 0.264 & 0.479 \\  
   \hline
\end{tabular}
\end{table}

\begin{table}[ht]
\centering
  \caption{NRMSE of estimating the x-coordinate deformation for varying number of windows and number of subsamples, normalized by the mean deformation of the x-coordinate}
   \label{t:2}
\begin{tabular}{r|rrrr}
\hline
  $\operatorname{NRMSE}_x$ &  \textbf{Subsamples}  \\ 
  \hline
  \textbf{Grid size }& 25 & 50 & 100 & 200 \\ 
  \hline
$3\times3$ & 0.612 & 0.430 & 0.279 & 0.112 \\ 
  $4\times4$ & 0.417 & 0.285 & 0.146 & 0.092 \\ 
  $5\times5$ & 0.341 & 0.203 & 0.139 & 0.074 \\ 
   \hline
\end{tabular}
\end{table}

\begin{table}[ht]
\centering
  \caption{RMSE and CRPS for the held-out testing set, comparing three registration methods}
   \label{t:rmse}
\begin{tabular}{rrrr}
  \hline
 & ICP & Rigid likelihood & Nonrigid likelihood \\ 
  \hline
RMSE & 0.086 & 0.118 & 0.080 \\ 
  CRPS & 0.044 & 0.027 & 0.021 \\ 
   \hline
\end{tabular}
\end{table}



\newpage


\bibliographystyle{MG}       
{\footnotesize
\bibliography{Chalk2.bib}}   

%

\end{document}